\documentclass[journal]{IEEEtran}
\usepackage{cite}

\ifCLASSINFOpdf
   \usepackage[pdftex]{graphicx}
\else
\fi
\usepackage{amsmath}
\usepackage{xcolor}

\begin{document}
%
\title{On The Structure of Plasma Jets in the\\ Rotating Plasma Experiment }
%
%
%

\author{V.~Valenzuela-Villaseca, 
L.~G.~Suttle,
F.~Suzuki-Vidal,
J.~W.~D.~Halliday, 
D.~R.~Russell,
S.~Merlini,
E.~R.~Tubman, 
J.~D.~Hare,
J.~P.~Chittenden,
M.~E.~Koepke, 
E.~G.~Blackman,  
S.~V.~Lebedev 

\thanks{V. Valenzuela-Villaseca was with the Department
of Physics, Imperial College London, UK. His current address is the Department of Astrophysical Sciences, Princeton University, USA. e-mail: v.valenzuela@princeton.edu.}
\thanks{L. G. Suttle, S. Merlini, E. R. Tubman, J. P. Chittenden, and S. V. Lebedev are with the Department
of Physics, Imperial College London, UK.}
\thanks{F. Suzuki-Vidal was with the Department
of Physics, Imperial College London, London,
SW7 2BW, UK. His current address is First Light Fusion Ltd., UK}
\thanks{J. W. D. Halliday was with the Department
of Physics, Imperial College London, London,
SW7 2BW, UK. His current address is at the Deparment of Physics, Clarendon Laboratory, University of Oxford, UK.}
\thanks{D. R. Russell was with the Department
of Physics, Imperial College London, London,
SW7 2BW, UK. His current address is at the Technical University of Munich}
\thanks{J. D. Hare is with the Plasma Science and Fusion Center, Massachusetts Institute of Technology, USA.}
\thanks{M. E. Koepke is with the Department of Physics at the University of West Virginia, USA.}
\thanks{E. G. Blackman is with the Department of Physics and Astronomy University of Rochester, USA.}


\thanks{Manuscript received MONTH NN, 2023; revised MONTH PP, 2024.}
}

%
%

\markboth{Journal of \LaTeX\ Class Files,~Vol.~14, No.~8, August~2015}%
{Shell \MakeLowercase{\textit{et al.}}: Bare Demo of IEEEtran.cls for IEEE Journals}
%



\maketitle

\begin{abstract}
Recent pulsed-power experiments have demonstrated the formation of astrophysically-relevant, differentially rotating plasmas \cite{Valenzuela-Villaseca2023}. Key features of the plasma flows are the discovery of a quasi-Keplerian rotation curve, the launching of highly-collimated angular-momentum-transporting axial jets, and a hollow density structure sustained by the centrifugal barrier effect. In this communication we discuss several features of the plasma structure in these experiments through order-of-magnitude models. First, we show that the observed rotation velocity would produce a centrifugal force strong enough to support the hollow density profile. Second, we show that the axial jet should diverge much faster than what was observed, were it not for a magnetized halo with 3 T which surrounds the jet and  exerts pressure on the interface. Finally, we discuss the temperature structure in the axial jet and plasma halo. We show that a $3$ T magnetic field would also suppress electron heat conduction, leading to the flat profile observed experimentally. We also find that the axial jet is efficiently radiatively cooled, whereas the halo is not, which would explain the thermal decoupling between the two regions.

\end{abstract}


%
\IEEEpeerreviewmaketitle

\section{Introduction}
Differentially rotating plasma flows are ubiquitous throughout the universe. They are commonly formed from gravitationally bounded flows and are typically structured as disks or tori oriented perpendicular to the angular momentum vector. A few examples of these systems are protoplanetary disks, black hole accretion disks, and spiral galaxies. Astronomical observations have identified that rotating flows are powerful jet engines on cosmic scales, launching streams of plasma along the rotation axis which often remain stable and collimated over extraordinary distances \cite{Begelman1984,Lynden-Bell1996,Blackman2020,Matsushita2021}.

Laboratory experiments can provide insights into the basic physics of differentially rotating plasma flows. In particular, they allow the measurement of the interplay between heating and cooling effects, density structure and rotation. In this communication, we discuss previously reported laboratory results from the Rotating Plasma Experiment (RPX) platform, a pulsed-power driven experiment designed to investigate key physics relevant to astrophysical accretion disks and jets \cite{Valenzuela-Villaseca2023,Bocchi2013,Ryutov2011, Valenzuela-Villaseca2022}. The central results were: 1) the platform generates a rotating plasma in which the angular momentum is dynamically significant and produces a hollow plasma density structure; 2) the absence of axial boundaries allows the ejection of axial outflows. These correspond to a central, cylindrical, dense, highly collimated jet surrounded by an unstructured halo; 3) the jet rotates and retains the hollow density profile, and; 4) the jet's ion temperature is significantly lower than in the halo. 

We characterize the system using simple calculations of, for example, energy budget and pressure balance. This paper is structured as follows: Section \ref{sec:setup} summarizes the experimental setup, diagnostics, and key experimental results. Section \ref{sec:density profile} discusses the observed density profile and quantifies the effect of angular momentum in the hydrodynamical equilibrium provided by the centrifugal barrier for an isothermal gas. Section \ref{sec:collimation} discusses the high degree of plasma collimation and stability. We hypothesize that the jet radial confinement must be provided by a strongly magnetized plasma halo. Section \ref{sec: temperature profile} uses the estimated magnetic field to show that electron heat transport is suppressed in the halo, allowing for an ion heat wave to diffuse outwards. We also investigate the effect of radiative cooling in the jet and halo regions and discuss why they have not thermalized. Finally, Section\ref{sec:conclusion} summarizes this work and suggests avenues for future investigation.

\section{Summary of Previous Experimental Setup and Results}\label{sec:setup}
\begin{figure}
    \centering
    \includegraphics[width=8.5cm]{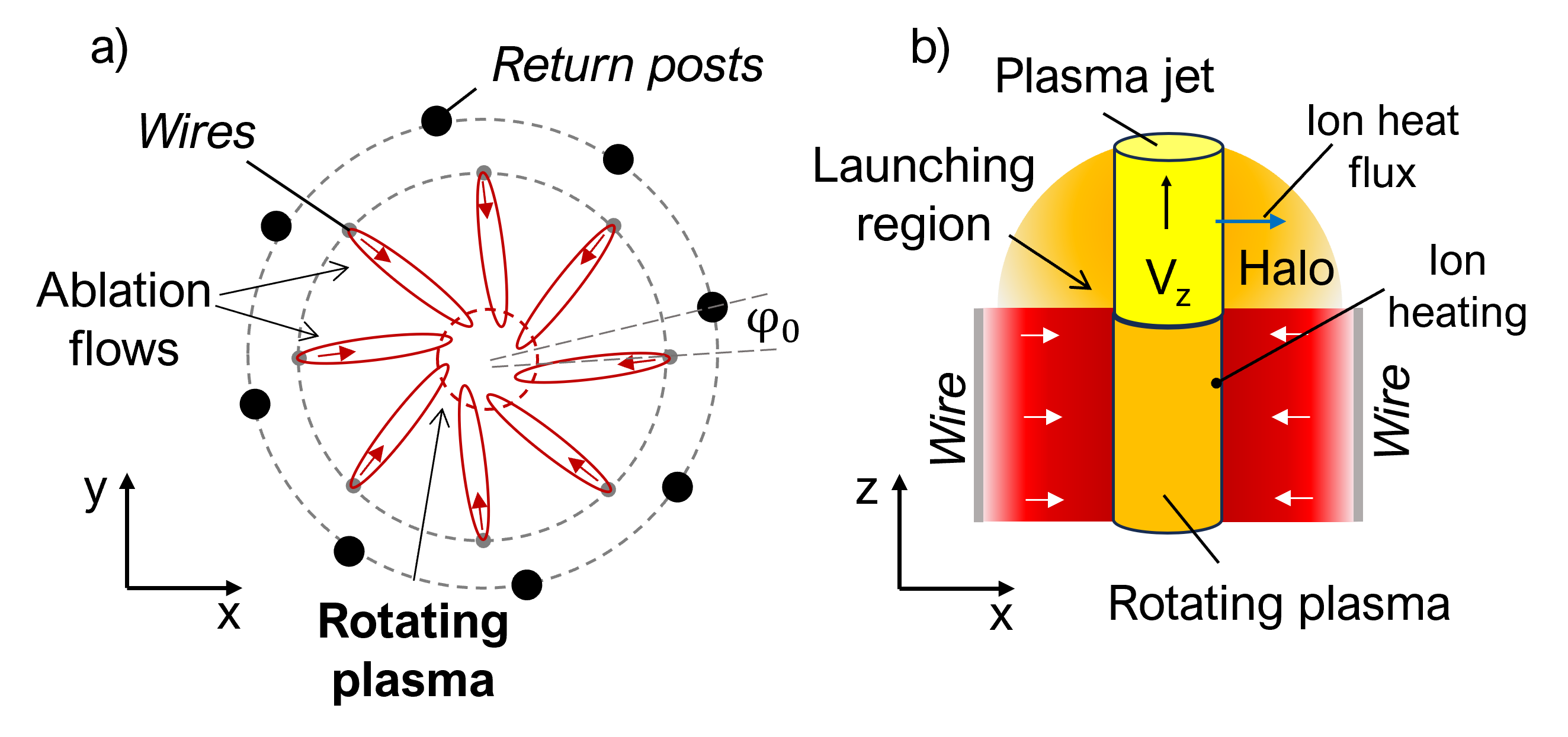}
    \caption{Schematic of the main experimental components. \textcolor{blue}{Eight aluminium wires (40 $\mu$m diameter each) are equally spaced on a 8 mm radius cylindrical configuration. The stainless steel return posts (1 mm diameter each) are located 11 mm from the axis. The off-set angle $\varphi_0=13^{\circ}$ }. a) End-on view. b) Side-on view. Relevant flow features are labelled.}
    \label{fig:schematic}
\end{figure}

RPX is a pulsed-power driven platform based on a cylindrical, ablating, wire array Z pinch \cite{Valenzuela-Villaseca2023}. The configuration was first proposed in numerical simulations by Bocchi et al.\cite{Bocchi2013} based on Ryutov's laser-driven concept \cite{Ryutov2011}. In the experiments, a wire array Z pinch was driven by the Mega-Ampere Generator for Plasma Implosion Experiments (MAGPIE) pulsed-power generator ($1.4$ MA peak current, $240$ ns rise-time) \cite{Mitchell1996}. Angular momentum is introduced in the ablation dynamics by enclosing the cylindrical wire array by an arrangement of return posts, with equal in number to the wires, with an overall off-set angle $\varphi_0$, shown in Figure \ref{fig:schematic}a.
The axial outflows (Figure \ref{fig:schematic}b) were probed using a combination of extreme-ultraviolet (XUV) pinhole imaging using micro-channel plate (MCP) detectors to gain qualitative information about the overall dynamics, Mach-Zehnder interferometry for imaging the electron density, and Thomson scattering (TS) for local measurements of temperature and flow velocity.

The key results from previous reports which are relevant for this communication are as follows:
\begin{itemize}
    \item[1. ]The plasma exhibits a density depletion towards the axis. This profile is consistent with the centrifugal barrier effect. This feature seems rather universal in experiments, as it has also been observed in a similar experiment investigated by Bennett et al. \cite{Bennett2015} (although the profile could not be unfolded from data), and numerically by Bocchi et al. \cite{Bocchi2013} in an RPX configuration.
    \item[2. ] The axial plasma expansion drives the formation of outflows, corresponding to a highly collimated axial jet with a full divergence angle $3^\circ$, and a low-density, unstructured plasma halo which surrounds the jet.
    \item[3. ] The plasma halo exhibits an unexpectedly high ion temperature $T_i = 250 \pm 50$ eV  near the edge of the jet (up to 2 mm from it). This is significantly higher than the typical ion temperature $T_i = 50 \pm 20$ eV observed inside the axial jet. At the same time, the electrons remain cool in the halo with $T_e < 15$ eV, whereas in the axial jet they have temperatures of $T_e = 30 \pm 10$ eV, with a steady increase towards the axis up to $T_e = 50 \pm 15$ eV.
\end{itemize}

Table I summarizes the plasma parameters of interest in the axial jet and halo regions. Detailed descriptions about methods and results can be found elsewhere \cite{Valenzuela-Villaseca2023,Valenzuela-Villaseca2022}.

\begin{table}
    \centering
\begin{tabular}{lccccc}
\hline\hline
\begin{tabular}[c]{@{}l@{}}Parameters\\ (units)\end{tabular}& & \begin{tabular}[c]{@{}c@{}}Axial jet\\ (edge)\end{tabular} & \begin{tabular}[c]{@{}c@{}}Axial jet\\ (core)\end{tabular} & \begin{tabular}[c]{@{}c@{}}Halo\\ (close)\end{tabular} & \begin{tabular}[c]{@{}c@{}}Halo\\ (far)\end{tabular} \\\hline
\begin{tabular}[c]{@{}l@{}}Radial location\\ (mm)\end{tabular}&$r$&$0.8$& $0$ & $1.2$ &$2.5$                               \\
\begin{tabular}[c]{@{}l@{}}Electron density\\ ($10^{18}$ cm$^{-3}$)\end{tabular}&$n_e$& \textcolor{blue}{$0.5$}  & \textcolor{blue}{$0.2$} &$0.2$    &$0.2$   \\
\begin{tabular}[c]{@{}l@{}}Max. rotation\\velocity\\ (km/s)\end{tabular}& $u_\theta$ & $23\pm3$ & $0$ & $0$ & $0$    \\
\begin{tabular}[c]{@{}l@{}}Max. radial\\velocity\\ (km/s)\end{tabular}&$u_r$& $2\pm3$ & $0$ & $-25\pm5$ & $-60\pm5$ \\
Charge state &$Z$& 7.2 & 6.8 & 3 & 3    \\
\begin{tabular}[c]{@{}l@{}}Electron\\temperature\\ (eV)\end{tabular}&$T_e$& $30\pm 10$ & $50 \pm 15$ & $< 15$ & $<15$ \\
\begin{tabular}[c]{@{}l@{}}Ion temperature\\ (eV)\end{tabular}&$T_i$& $50\pm15$ &$70\pm20$ & $250\pm50$ & $40\pm10$   \\
\hline\hline \\
\end{tabular}
    \caption{Summary of plasma parameters relevant to the calculations reported in \cite{Valenzuela-Villaseca2023}. \textcolor{blue}{Measurements were made across both types of outflows at $z=5.5$ mm above the upper end of the wires (see Fig. 1b)}. The ion-acoustic wave feature is single peaked in the halo region, and therefore the electron temperature has only an upper constraint.}
    \label{tab:parameter summary}
\end{table}

\section{Density Structure and the Centrifugal Barrier Effect}\label{sec:density profile}

The differentially rotating plasma flow exhibits two important properties regarding its internal energy. The plasma is iso-thermal, with an approximately flat temperature distribution in the bulk. The plasma also exhibits a steep density profile with a density depletion close to the axis. In the inviscid, unmagnetized case, this equilibrium corresponds to solutions to the axisymmetric hydrodynamical radial equation
\begin{equation}
    \frac{dp}{dr} = \rho r \Omega^2,
\end{equation}
where $p$ is the fluid's internal pressure, $\rho$ is the mass density, $r$ is the radial coordinate, and $\Omega = \Omega(r)$ is the flow's angular frequency. Assuming an ideal gas equation of state $p = k_B n_i(T_i + ZT_e)$, where $k_B$ is Boltzmann's constant, $n_i$ the ion number density, $T_i$ the ion temperature, $Z$ the average charge state, and $T_e$ the electron temperature, and imposing an isothermal distribution, the centrifugal barrier is supported solely by a density gradient given by the static equation
\begin{equation}\label{eq:balance}
    \frac{1}{n_i}\frac{dn_i}{dr} = \frac{r\Omega^2}{c_s^2},
\end{equation}
where $c_s \equiv \sqrt{k_B(T_i+ZT_e)/mi}$ is the ion-acoustic speed $c_s$ and $m_i$ the ion mass. Equation (\ref{eq:balance}) is a solvable model for the density profile, given a rotation curve parameterized by $\Omega = \Omega(r)$. Let us consider the simplest case of rigid-like rotator such that $\Omega = \Omega_0 =$ constant. This case may be relevant to a high-$\beta$, viscous plasma such as the Plasma Couette Experiment \cite{Collins2012}, were an external electromagnetic torque is applied at the boundary and the viscous stress imprints angular momentum on the bulk. If at the outer boundary $r = r_0$ the density is $n_i(r=r_0) = n_0$, then one finds
\begin{equation}\label{eq:rigid}
    n_i(r) = n_0 \exp{\left[\frac{(r^2-r_0^2)\Omega_0^2}{2c_s^2}\right]}.
\end{equation}
Notice that the shape of the density profile is determined by the Mach number M$=r\Omega/c_s$. Therefore, if rotation is subsonic, the fluid is essentially incompressible and a shallow profile is enough to sustain the centrifugal force, as assumed in the Taylor-Couette problem \cite{Chandrasekhar1961}.

A more complicated (yet physical) rotation curve can be Taylor-expanded in power-law terms such that $\Omega = A r^\lambda$ where $A$ is a dimensionally-appropriate constant. We notice that the Keplerian case has $\lambda = -3/2$. For a general power-law such that $\lambda \neq -1$, the density profile is given by
\begin{equation}
    n_i(r) = n_0 \exp\left[B(\lambda) \left( r^{2(\lambda+1)} - r_0^{2(\lambda+1)} \right)  \right],
\end{equation}
where $B(\lambda) \equiv A^2 /2(\lambda+1)$. The case $\lambda = 0 \Leftrightarrow \Omega = \Omega_0=$ constant yields the solution (\ref{eq:rigid}).  

The case $\lambda = -1$ corresponds to constant velocity $u_0 \equiv r\Omega = $ constant, the integration needs to be done at an arbitrarily small cut-off radius $r_{min} > 0$ so that $\Omega$ is well-defined in the domain. Solving for the density yields the profile
\begin{equation}
    n_i(r) = n_0 \left( \frac{r}{r_0} \right)^{\frac{u_0^2}{c_s^2}},
\end{equation}
where the exponent is again determined by the Mach number M $= u_0/c_s$.

Any continuously differentiable density profile can be written as
\begin{equation}
    \frac{n_i(r)}{n_0} = \left( \frac{r}{r_0} \right)^{\frac{u_0^2}{c_s^2}} + \sum_{j\neq -1} \exp\left[B_j \left( r^{2(\lambda_j+1)} - r_0^{2(\lambda_j+1)} \right)  \right], 
\end{equation}
where $B_j \equiv B(\lambda_j,A_j)$, for a set of exponents $\{\lambda_j\}$, under the prescription $r > 0$. 

\begin{figure}
    \centering
    \includegraphics[width=7.5cm]{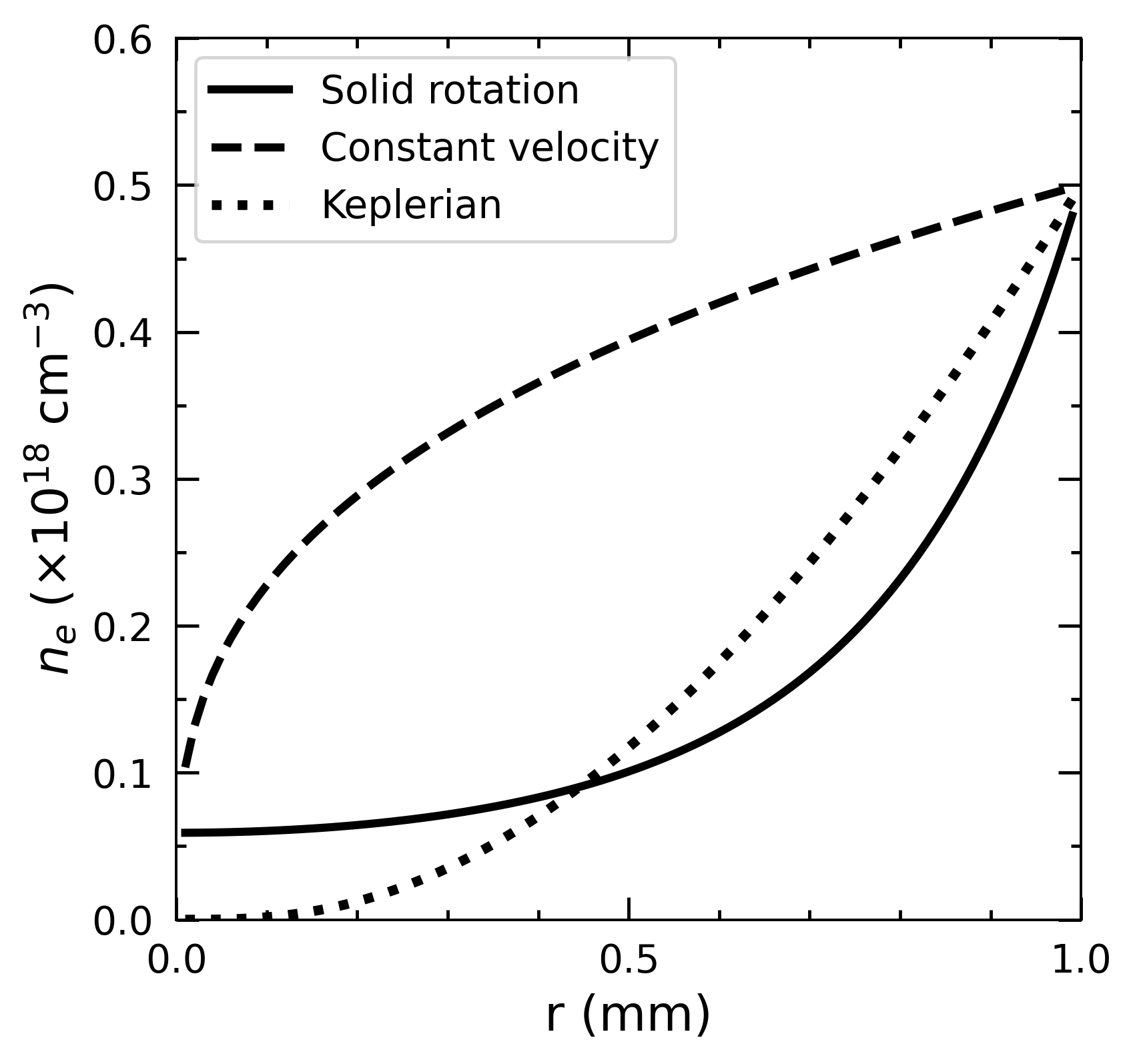}
    \caption{Examples of calculated \textcolor{blue}{electron} density profiles \textcolor{blue}{(assuming quasi-neutrality $n_i = Z n_e$)} with $n_0 = 5\times 10^{17}$ cm$^{-3}$, $T_e = 40$ eV, $T_e= 50$ eV, $Z = 7$. The rigid solid case (solid line) was calculated for $\Omega_0 = 5\times 10^{7}$ rad/s. \textcolor{blue}{Constant velocity} (dashed line) was calculated using $u_0 = 20$ km/s. The power-law (dotted line) corresponds to Keplerian rotation, i.e. $\lambda = -3/2$ and $A=0.7$. }
    \label{fig:density profiles}
\end{figure}

Figure \ref{fig:density profiles} shows a few examples \textcolor{blue}{of electron density profiles $n_e = n_e(r)$ calculated using the quasi-neutrality condition because this is what is measured experimentally}. Notice that in the regime of interest, all of the profiles are steep enough to be consistent with the observed data. However, it is notable that the curve inflexion is different. A concave curve corresponds to supersonic rotation M $>1$, whereas convex curves correspond to the subsonic case M $<1$.

This model is valid only for iso-thermal flows, and is therefore reasonable when the thermal dissipation length is comparable to the plasma radius. In these conditions, it may be a reasonable approximation which would \textcolor{blue}{allow} estimating the rotation velocity from the density profile. In previous reports, either Thomson Scattering (e.g. \cite{Valenzuela-Villaseca2023,Bennett2015}) or Doppler shift self-emission spectroscopy (e.g. \cite{Cjevic2022}) has been utilize to diagnose rotation velocity. In the absence of such diagnostics, it may be possible to still infer the rotation velocity using imaging diagnostics instead. If the electron-ion equilibration time is short compared to the hydrodynamic timescales, and radiatively cooling is negligible, then $T_i \approx T_e$ and it is possible to use X-ray detector, such as filtered Photo-cathode diodes (PCD) or micro-channel plate (MCP) detectors, to accurately infer the plasma temperature and therefore the ion-acoustic sounds speed, since in this regime $c_s \propto \sqrt{T_e}$. By coupling those measurements with laser interferometry to infer the electron density profile, a rotation velocity could be inferred. 

%


\section{Jet Collimation by a Magnetized Halo}\label{sec:collimation}
Identifying the mechanism for collimation of astrophysical jets is an outstanding challenge. Theoretical models employ inertial or magnetic confinement effects to explain the high aspect ratio of these structures. In particular, magnetic collimation seems to be a natural mechanism, however it can become subject to plasma instabilities that may disrupt the jet \cite{Huarte-Espinosa2012}. 

Intuitively, we expect differentially rotating, laboratory plasma jets to have poor collimation. They are hot, dense, and spinning. Therefore, the internal pressure pushes outwards strongly (or at least more strongly than the non-spinning case), adding stress to any confinement mechanism. In fact, numerical simulations by Bocchi et al. \cite{Bocchi2013} show divergence angles $>15^{\circ}$. On the contrary, RPX jets are remarkably  collimated, with divergence angles $\sim 3^{\circ}$, while also remaining stable up until $\sim 270$ ns.

We consider two cases, non-rotating and rotating, unmagnetized plasma jets propagating in a vacuum at speed $V_z$. In the non-rotating case, the full divergence angle $\phi_{NR}$ is given by
\begin{equation}
\phi_{NR} = 2\tan^{-1}\left(\frac{c_s}{V_z} \right).    
\end{equation}
By inspection we find that the divergence angle decreases with increasing sonic Mach number M $=V_z/c_s$. Taking edge numbers of the jet from Table \ref{tab:parameter summary}, we calculate $\phi_{NR} = 37^{\circ}$, which is one order of magnitude above the observed angle. 

Working in the co-rotating frame of reference, the centrifugal force pushes outwards with pressure $p_c = \rho u_\theta^2$. To calculate the effective Mach number, we can define a centrifugal temperature $T_c = p_c/k_Bn_i$, such that it defines an equivalent sound speed $c_{equiv} = (k_B T_c / m_i)^{1/2} = u_\theta$. Therefore, the effective Mach number which includes rotation is given by
\begin{equation}
    \text{M}_{\text{eff}} = \frac{V_z}{c_s+ u_\theta}.
\end{equation}
This shows (maybe obviously, in retrospect) that the rotation velocity enters in the definition of the Mach number exactly as a sound speed. Defining the divergence angle as

\begin{equation}
    \phi_R = 2\tan^{-1}\left( \frac{1}{\text{M}_{\text{eff}}} \right),
\end{equation}

and taking $u_\theta = 20$ km/s, we find $\phi_R = 50^{\circ}$. This shows that the presence of a dynamically significant angular momentum greatly impacts the collimation of these plasma jets. 

Let us now drop the assumption that the jet is propagating through a vacuum and instead consider the effect of the halo in providing extra confinement. Thomson scattering Doppler shift indicates that the halo is inflowing, and therefore providing an inwards ram pressure. Moreover, the ion temperature reaches up to $250$ eV in the near vicinity of the jet, which would contribute additional thermal pressure. If we assume perfect collimation and that jet and halo move upwards at comparable speed, at the interface, a pressure balance equation can be written
\begin{equation}\label{eq:pressure balance}
    p_{th,jet} + p_{c,jet} = p_{th,halo} + p_{ram,halo} + p_{mag,halo},
\end{equation}
where the left hand side has the thermal and centrifugal pressure of the jet, respectively, and the right hand side the thermal, ram, and magnetic pressure of the halo, respectively.

\begin{figure}
    \centering
    \includegraphics[width=7.5cm]{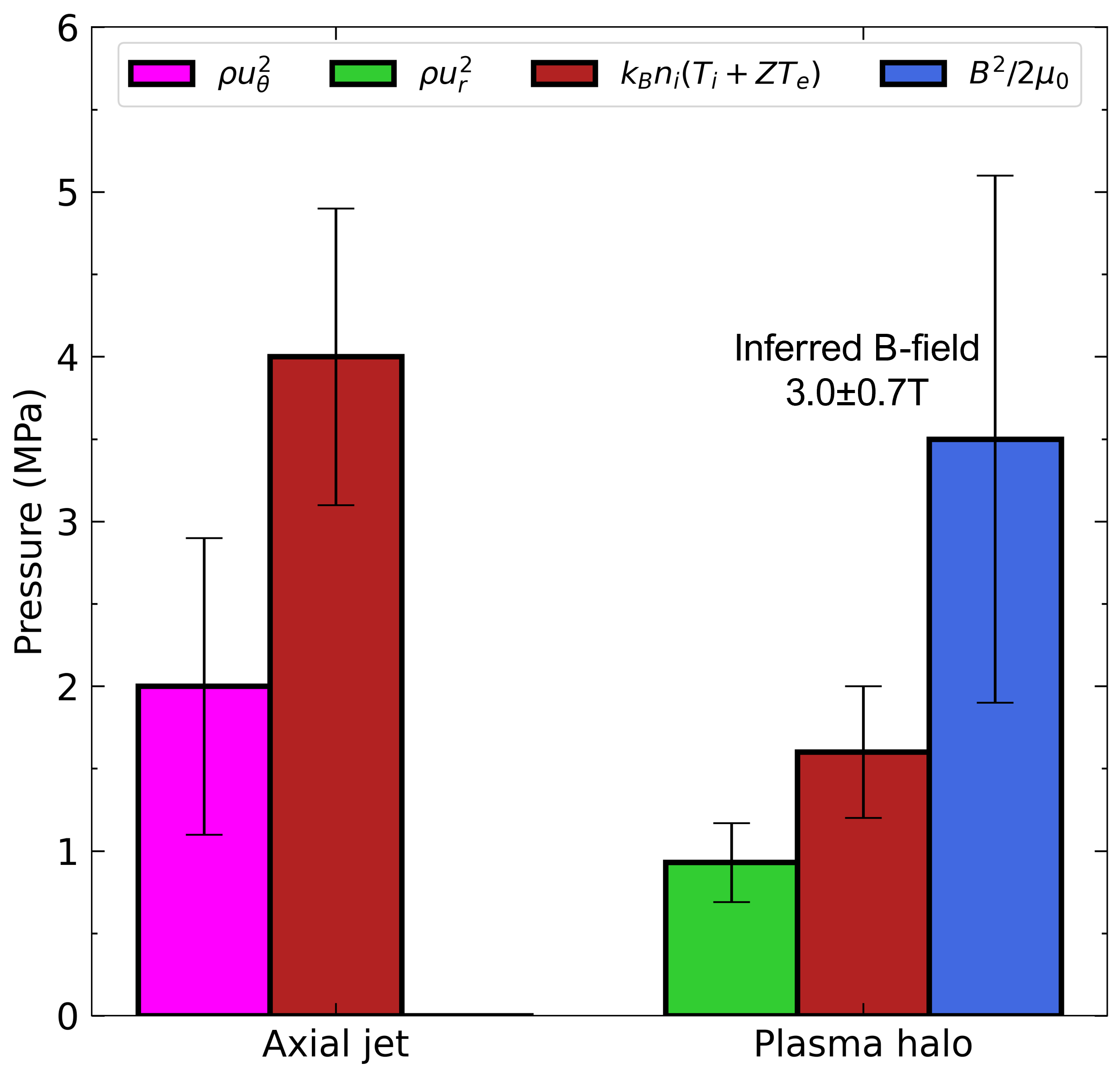}
    \caption{Partial calculated pressures for the axial jet and plasma halo.}
    \label{fig:pressure balance}
\end{figure}

Figure \ref{fig:pressure balance} shows the relative contributions of each component from the numbers shown in Table \ref{tab:parameter summary}. The bar graph shows that the outwards pushing thermal and centrifugal pressure are stronger than their counterparts in the halo, and therefore a magnetic pressure needs to be added which corresponds to an inferred $B = 3.0 \pm 0.7$ T magnetic field. The Figure also shows that the thermal-$\beta$ (thermal-to-magnetic pressure ratio) is $\beta \sim 0.5$.

There is more evidence of this magnetic field in the halo. Thomson scattering measurements indicate that the halo is flowing inwards. At large radial distances $r > 2.2$ mm, the plasma reaches a maximum radial velocity of $u_r = -60$ km/s, which is consistent with the inwards acceleration of ablation flows in wire array Z pinches \cite{Harvey-Thompson2012}. However, at $r=2.2$ mm, the plasma decelerates to a velocity of $-25$ km/s before reaching the jet, and in the absence of a density jump. Balancing the change in ram pressure with an increase in magnetic pressure over the same length-scale, we estimate a characteristic $B=2.7\pm 0.3$ T, i.e. of the same order as inferred from jet collimation estimate.

In this analysis we have neglected the magnetization of the jet. A magnetic field frozen-in the jet would increase the outwards pressure but adding hoop stress via magnetic tension that can contribute to the jet confinement. In a Bennett pinch, these forces are balanced \cite{Friedberg1987}. Rotation adds a caveat, however. In the presence of an initial radial field \textcolor{blue}{$B_r$}, the azimuthal magnetic field \textcolor{blue}{$B_\theta$} can grow linearly in time due to the $\omega$-effect \cite{Verhille2010}. In the ideal limit, this radial field may be expected because it is what introduces the initial angular momentum. \textcolor{blue}{The linearly growing magnetic field strength is given by \cite{Balbus1991}}
\begin{equation}
    \textcolor{blue}{|B_\theta(t)| = \left|rB_r \frac{d}{dr}\left(\frac{u_\theta}{r} \right) \right| t,}
\end{equation}
\textcolor{blue}{where to first order approximation $u_\theta \sim$ constant, and therefore}
\begin{equation}
    \textcolor{blue}{|B_\theta(t)| \sim \frac{u_\theta B_r}{r}t.}
\end{equation}
\textcolor{blue}{Let us assume that all the radial field from the wire ablation is advected and take $B_r = 2$ T \cite{Valenzuela-Villaseca2023}, $u_\theta = 20$ km/s, and evaluate at the plasma half-radius $r=0.5$ mm. Then in $100$ ns, $B_\theta = 8$ T, which is comparable to the magnetic field ablating the wire array Z-pinch.} 

As the magnetic field grows linearly, both the magnetic pressure and tension grow quadratically in time. However, the thermal pressure does not. Therefore, \textcolor{blue}{in} an initially equilibrated rotating plasma, with initial Bennett equilibrium condition $\beta = 1$, would break as the $\omega$-effect forces $\beta$ to fall below unity, breaking the equilibrium and forcing the plasma to pinch. \textcolor{blue}{In Bennett equilibrium, the total outwards fluid pressure is given by the left-hand side of equation (\ref{eq:pressure balance}) $p_{total,jet} = p_{th,jet} + p_{c,jet} = 6$ MPa, for which a $B_{\theta,\text{Bennett}} = 4.1$ T is sufficient to keep $\beta$ equal to unity. Therefore, in the presence of the $\omega$-effect, the plasma would violently pinch on axis.} This is not what is observed in the experiment, \textcolor{blue}{where the jet} remains in steady state. Hence, we propose an alternative steady configuration consisting of an unmagnetized jet, which remains confined and stable \textcolor{blue}{by} the combination of thermal, ram, and magnetic pressure provided by the halo. \textcolor{blue}{This would be supplied by a skin current flowing through the outer edge of the jet and returning through the halo, as previously suggested by \cite{Lebedev2005}. Quantitatively, a $10 - 20$ kA skin current is enough to supply a $3$ T field in the jet's vecinity, which is consistent with previous measurements on ablating wire array Z-pinches \cite{Bott2006}}

\section{Temperature Profile and Heat Transport Through the Halo}\label{sec: temperature profile}

We now discuss the temperature profiles in the jet and halo. The main results show that the electron temperature increases towards the axis and is minimum in the halo. On the contrary, ion temperature is maximal in the near vicinity of the jet. In this section we address three questions:
\begin{itemize}
    \item[1.] Why is the ion temperature higher in the halo ($T_i \sim 250$ eV) than in the jet ($T_i \sim 50$ eV)?
    \item[2.] Why is the electron temperature in the halo low ($T_e < 15$ eV) even near the jet?
    \item[3.] Why the electrons and ions remain thermally unequilibrated in the jet? 
\end{itemize}

The increasing ion temperature does not occur in a region of increasing density, therefore it cannot be explained as a form of compression heating. Alternatively, an electrical current may be circulating in the jet's vicinity, however that would increase the electron temperature through Joule heating instead. 

In a typical high energy density experiment, ion heating can only come from compression (or shock heating). The only region in the experiment where that can take place is inside the wire array, where the ablation flows merge with the central rotating plasma column. In a similar experiment, Bennett reported ion temperatures of up to $T_i \sim 600$ eV near the outer rim of the rotating plasma \cite{Bennett2015b}. We therefore conclude that the origin of the increase ion temperature must be in that interaction region.

We hypothesize that \textcolor{blue}{the primary source of plasma heating after ablation is due to shock compression in the region where the ablation flows merge with the rotating plasma. This will initially heat up the ions rather than the electrons, which are expected to maintain a higher temperature than the radiatively cooled electrons due to the long energy equilibration timescale and short radiative cooling timescale, with respect to the dynamics of interaction (we quantify this, in detail below). As the plasma expands axially and the jet is launched upwards, it propagates into the halo, launching a thermal diffusion wave outwards. The precise shape of the temperature profile inside and outside of the jet must therefore be determined by competing heating, cooling, and transport mechanisms.}

Since the halo is flowing into a density jump, we will think the problem to be similar to a thermal diffusive shock \cite{Drake2018,Russell2021}. Therefore, the temperature profile can be described in terms of the thermal diffusion length and time scales, which need to be computed from transport coefficients.

The heat diffusion coefficients depend on the particle magnetization \cite{Braginskii1965} described by the Hall parameter $\tau_{ee}\Omega_{c,e}$ and $\tau_{i}\Omega_{c,i}$ for electrons and ions, respectively, where
\begin{align}\label{eq:col}
    \tau_{ee} = \frac{3.5\times 10^5}{\ln(\Lambda)}\frac{T_e^{3/2}}{Zn_e}, \tau_{ii} = \frac{3\times 10^7}{\ln(\Lambda)}\sqrt{ \frac{m_i}{2m_p}}\frac{T_i^{3/2}}{Z^3n_e},
\end{align}
are the electron-electron and ion-ion collision times ($\ln(\Lambda)$ is the Coulomb logarithm, $m_i$ is the ion mass, and $m_p$ is the proton mass), and 
\begin{align}\label{eq:gyro}
    \Omega_{c,e} = \frac{m_e}{eB},\quad \Omega_{c,i} = \frac{m_i}{ZeB},
\end{align}
and are the electron and ion cyclotron frequencies ($m_e$ is the electron mass, and $e$ is the fundamental charge). Equation (\ref{eq:col}) above is for temperatures in eV and densities in cm$^-3$.

When the Hall parameter is below unity, binary Coulomb collisions prevent the particles from completing orbits around the magnetic field lines. On the contrary, if the parameter is greater than unity, the particles can (on average) complete gyro-orbits before undergoing a collision, and hence are called magnetized. In this case, the net effect of the magnetic field is to decrease particle mobility, suppressing heat transport. Notice that the particles always gyrate in a plane perpendicular to the local magnetic field direction. Therefore, this suppression occurs only in the perpendicular direction. This leads to anisotropic transport coefficients where the particles can stream along the magnetic field as if they were unmagnetized. Henceforth, we will use the symbol $\parallel$ to denote transport coefficients in cases where either in the unmagnetized or parallel to the magnetic field, and the symbol $\perp$ for transport coefficients perpendicular to the magnetic field. Assuming a $3$ T field in the halo, the Hall parameter for the electrons is $\sim 5$ and ions $\sim 0.1$. Therefore, the electrons are magnetized whereas the ions are not.

Let us imagine that an ion heat diffusion wave propagates outwards from the jet and into the halo. The length scale of the wave is approximately 1 mm. Let us calculate the unmagnetized ion thermal diffusion length given by
\begin{equation}
    \ell_{i,\parallel} = \frac{\chi_{i,\parallel}}{u_r},
\end{equation}
where the ion thermal diffusion coefficient is given by
\begin{equation}
    \chi_{i,\parallel} = 3.9 \frac{\tau_{ii}}{m_i}T_i.
\end{equation}
Taking the inflow velocity near the jet $u_r = 25$ km/s yields $\ell_{i,\parallel} = 1.2$ mm, which is approximately what we see in the experiment. The same argument can be used to show that the electron magnetization suppresses the heat transport. Their diffusion coefficient is given by
\begin{equation}
    \chi_{e,\perp} = 4.7 \frac{T_e}{m_e \Omega_{c,e}^2\tau_{ee}},
\end{equation}
which yields a thermal diffusion length for magnetized electrons to be $\ell_{i,\perp} = 100$ $\mu$m. Therefore, an azimuthal magnetic field would prevent the electrons from conducting heat from the jet into the halo. Nevertheless, as the ions in the halo heat up, they can exchange energy with the electrons leading towards equilibration. However, the electron-ion equilibration time $\tau^{e\backslash i}_{eq} \sim 60$ ns \cite{Huba2016}, which is longer than the transit time $t_{transit} \sim 25$ ns of the inflow through the hot halo. Therefore, the electrons cannot heat up efficiently through thermal diffusion from the jet nor through binary collisions with the ions. 

The measurements on the RPX experiments were done at a characteristic height $H\sim 6$ mm above the upper end of the launching region. We can ask the question: once the jet is launched upwards, what is the thermal diffusion timescale and how does it compare to the jet transit time to reach the probing region? In other words, is the diffusion time scale consistent with the time of flight of the jet? The thermal diffusion time is given by \textcolor{blue}{$\tau_{diff} = \ell_{i,\parallel}^2/\chi_{i,\parallel} \sim 65$ ns}, and corresponds to the timescale over which the heat wave can propagate one diffusion lengthscale. The jet was estimated to have a vertical velocity $V_z \sim 100$ km/s, and therefore the time of flight is given by $t_{f} = H/V_z = 60$ ns $\approx \tau_{diff}$. This shows that the wave is expected to travel approximately 1 mm outwards from the surface of the jet, once the jet has been ejected upwards.

\begin{figure}
    \centering
    \includegraphics[width=7.5cm]{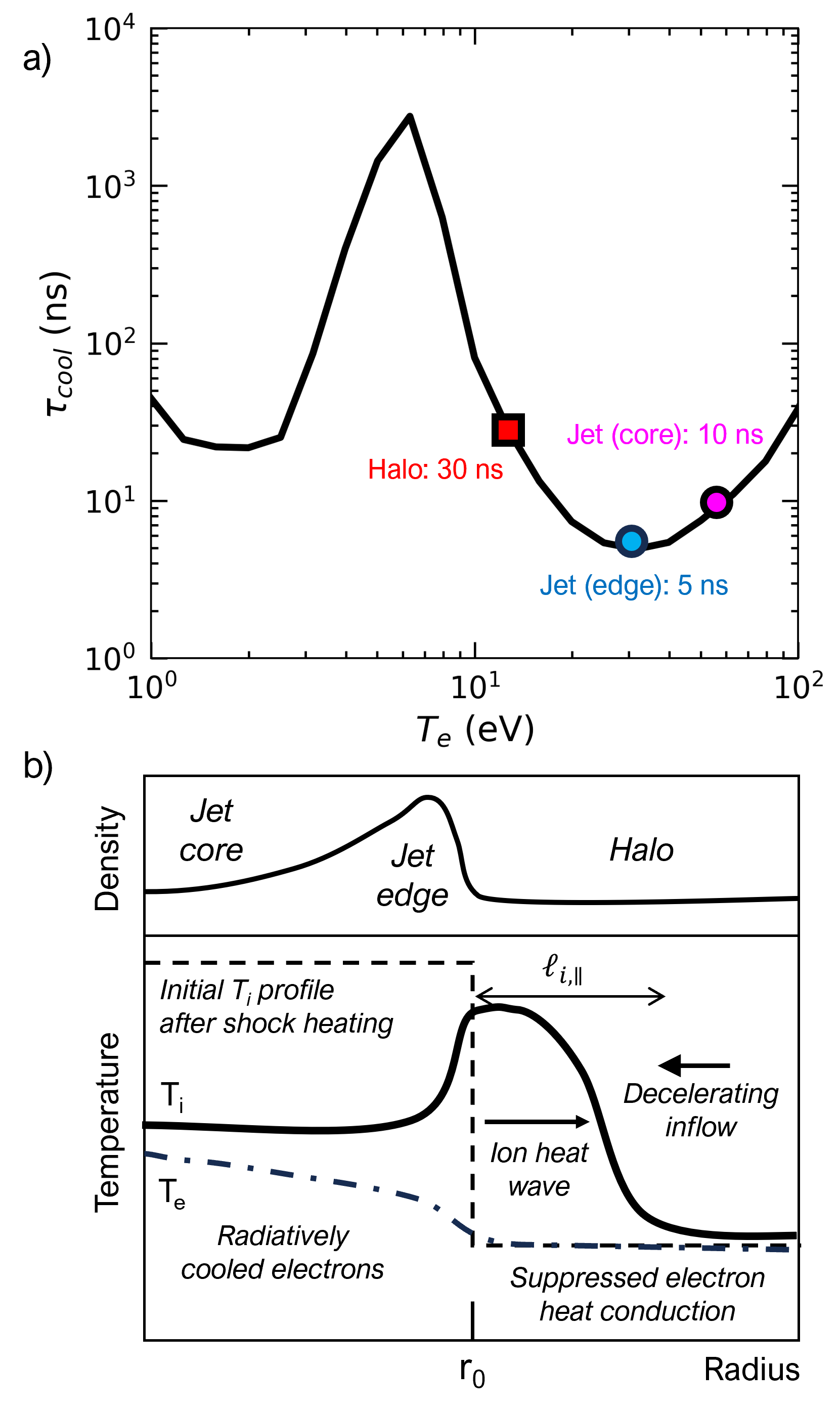}
    \caption{a) Cooling time as a function of electron temperature for aluminium and $n_i = 10^{17}$ cm$^{-3}$ \cite{Russell2021,Suzuki-Vidal2015}. Estimated values for jet and halo conditions are annotated. b) Expected temperature profile from an initial step-like temperature profile (dotted line). The plasma radius $r_0$ determines the separation between jet and halo, i.e the jet is the region such that $r<r_0$.\textcolor{blue}{The halo is inferred to be magnetized by a 3 T field, whereas the jet remain unmagnetized. b) Sketch of the hypothesized temperature evolution. Upper panel: density profile. Bottom panel: Initial ion temperature profile after shock heating. At later times, an ion heat wave propagates into the halo increasing ion temperature (solid line) at $\sim 1 \ell_{i,\parallel}$ from the edge. At the same time, electron heat flux is mitigated by the (estimated) azimuthal magnetic field, and therefore the electron temperature (dotted dashed line) remains flat in the halo. In addition, at the core, the radiatively cooled electrons can exchange energy with the ions efficiently, thus cooling down both species. Both electron-ion equilibration and radiative cooling are more efficient at the edge than the core.} }
    \label{fig:cooling time}
\end{figure}

We now focus on the inside of the jet, which has lower temperature than the halo. Since the jet is more dense than the halo, the electron-ion equilibration time is $\sim 15$ ns, significantly faster. However, in an aluminium plasma, we expect radiative cooling to become significant at these densities. The radiative cooling time is given by \cite{Russell2021,Suzuki-Vidal2015}
\begin{equation}
    \tau_{cool} = 2.4\times 10^{-12} \frac{(1+Z)T_e}{Z n_i \Lambda(n_i,T_e)},
\end{equation}
where $\Lambda(n_i,T_e)$ is the cooling function (not to be confused with the argument in the Coulomb logarithm).

Figure \ref{fig:cooling time}a shows the cooling time for an aluminium plasma in the relevant conditions. Available simulation results from \cite{Suzuki-Vidal2015,Russell2021} had slightly higher densities than what we measured experimentally, but we do not expect large deviations. They show that the halo is inefficiently cooled. However, the jet exhibits strong cooling, especially at the edge. This is consistent with colder electrons on the outer regions of the jet where the density is larger, but a steady increase towards the axis where density is depleted. Since, in the jet $\tau_{cool} < \tau^{e\backslash i}_{eq}$, then we expect some thermal decoupling from the electrons and the ions. Moreover since both $< t_{f}$, we expect that the electrons gain energy from the ions, but also radiate energy efficiently. Hence, the ions get cooled indirectly via energy exchange with the electrons, which then radiate energy away.

The resulting temperature profile from these processes is shown schematically in Figure \ref{fig:cooling time}b. The top panel shows the density profile, with the jet described as the dense shell surrounding a depleted axis. The halo is the region $r>r_0$. An initial, step-like temperature profile is imprinted from the interaction of the ablation flows and the wire array. As the jet is ejected, a thermal diffusion wave heats up the ions in the halo up to a difussion length $\ell_{i,\parallel}$, however the magnetic field suppresses electron transport. At the same time, electrons and ions can exchange energy efficiently inside the jet region. However, radiative cooling prevents full equilibriation. The density profile produces slightly more inefficient cooling near the axis, which produces a temperature increase towards the axis. Notice that this interpretation is completely consistent with the, independently inferred, $3$ T magnetic field in the halo.

\textcolor{blue}{Interestingly, similar temperature profiles have been found in AGN accretion disk coronae, large low density regions above and below the disk. In such environments, electrons cool via Compton scattering leading to a two-temperature magnetized plasma where main thermal conduction is provided by the ions \cite{Fabian2015,Bambic2024}. Thus, RPX may provide a platform over which study heat transport and radiative cooling effects relevant to such environments.}

\section{Conclusion}\label{sec:conclusion}

We have discussed important features of the experiments reported in \cite{Valenzuela-Villaseca2023}. The main conclusions are

\begin{itemize}
    \item[1.] The observed density profile is consistent with an unmagnetized, iso-thermal centrifugal barrier. By writing a balance equation we have shown that the shape of the density profile is determined by the rotation curve. We suggest that this result could be used to indirectly infer the plasma rotation curve in the absence of spectroscopic techniques.
    \item[2.] The jet is likely collimated by a strongly magnetized halo. The data seems to suggest that there is a $B=3$ T magnetic field in the vicinity of the jet, which in addition to the inwards ram pressure and thermal pressure provided by the ions, seem to confine radially the jet. Additional evidence comes from the deceleration of the inflowing halo, which can also be explained by a magnetic field of the same order. This structure may be of interest for the astrophysics community, since observations of active galactic nuclei (AGNs) have shown the formation of so-called coronae which are expected to be strongly magnetized. Therefore, RPX could be an ideal platform to investigate the fundamental interactions of rotating jets launched by black hole accretion disks with highly magnetized, diluted plasmas.
    \item[3.] The thermal structure of the plasma is likely due to the combined effect of radiative cooling in the jet, which is reduced towards the axis as the density is lower there, and suppression of electron heat conduction through the halo. It is likely that the high ion temperatures in the halo are the result of a thermal diffusion wave launched from the edge of the jet and generated by the interaction of the ablation flows and the rotating plasma in the wire array region. 
\end{itemize}

\textcolor{blue}{A general conclusion is that RPX jets exhibit an dynamically significant interplay between plasma properties. Rotation sets a density profile capable of controlling the radiative cooling rate. In turn, the plasma temperature regulates the steepness of the density gradient sustained by the centrifugal barrier effect, both of which affect the cooling rate. It remains an open question if in a longer-lived experiment can be used to test if these effects can trigger and sustain new interesting radiation-MHD equilibria and/or instabilities in rotating plasmas.}

Future work will be to probe directly the magnetic field in the halo and jet to compare with the models presented in this paper and try different wire materials to investigate the impact of weaker and stronger radiative cooling effects.


%



\section*{Acknowledgment}

This work was supported in part by NNSA under DOE Cooperative Agreement No DE-SC0020434 and DE-NA0003764. Vicente Valenzuela-Villaseca was funded by the Imperial College President's PhD Scholarships and the Royal Astronomical Society. We thank Chris J. Bambic for teaching us about accretion disk coronae and enthusiastic conversations about their potential relation to our experiments.

\ifCLASSOPTIONcaptionsoff
  \newpage
\fi



%

%

\begin{IEEEbiographynophoto}{Vicente Valenzuela-Villaseca}
is a Postdoctoral Research Associate at the Department of Astrophysical Sciences at Princeton University (USA). He completed his undergraduate and masters degrees in Experimental Physics at Pontificia Universidad Catolica de Chile, and received his PhD in Plasma Physics at Imperial College London in 2022. he has led experimental campaigns on the MAGPIE pulsed-power generator (UK) and the OMEGA laser system (USA), focusing on High-Energy-Density Laboratory Astrophysics experiments. 
\end{IEEEbiographynophoto}

\begin{IEEEbiographynophoto}{Lee George Suttle}
is a Postdoctoral Research Associate in the Plasma Physics Group at Imperial College London, UK. His research interests focus on magnetized high energy density phenomena (shocks, reconnection, instabilities) and high temperature plasma diagnostic development.
\end{IEEEbiographynophoto}

\begin{IEEEbiographynophoto}{Francisco Suzuki-Vidal}
is a plasma physics researcher in High Energy Density Physics and Inertial Confinement Fusion. He obtained his PhD at Imperial College London in 2009 where he led experiments on the MAGPIE pulsed-power facility investigating the formation of magnetically-driven plasma jets. He was then awarded a Fellowship from The Royal Society which allowed him to lead experiments at the PALS, Orion, SG-II and OMEGA laser facilities. He is currently Lead Scientist at First Light Fusion looking at engaging with the international plasma and ICF communities through external collaborations.
\end{IEEEbiographynophoto}

\begin{IEEEbiographynophoto}{Jack W. D. Halliday,}
photograph and biography are not available at the time of publication.
\end{IEEEbiographynophoto}

\begin{IEEEbiographynophoto}{Danny R. Russell}
completed his undergraduate studies at Imperial College London. He did a PhD in the plasma physics group at Imperial studying plasma shocks followed by a postdoctoral position in the same group. He is currently a postdoctoral researcher at the Technical University of Munich.
\end{IEEEbiographynophoto}

\begin{IEEEbiographynophoto}{Stefano Merlini}
is a Research Assistant in the Plasma Physics group at Imperial College London. He received a BSc in Bioengineering and a double Master’s degree in Aerospace Engineering from University of Padova (Italy) and Cranfield University (UK). Stefano’s research field focuses on shocks and transition to turbulence in supersonic magnetised HED plasmas relevant to astrophysics and Inertial Confinement Fusion. During his PhD, he has conducted experiments on inverse Z-Pinch exploding wire arrays at the MAGPIE pulsed-power facility using laser-based diagnostics, such as Thomson Scattering, laser interferometry and a novel diagnostic technique, named imaging refractometer, allowing to characterise electron density perturbations in high-dense plasmas
\end{IEEEbiographynophoto}

\begin{IEEEbiographynophoto}{Eleanor Rose Tubman}
is a lecturer at Imperial College London in experimental high energy density plasma physics. Her primary area of expertise lies in understanding the influence of magnetic fields in laser-produced plasmas. She received her PhD from the University of York in 2012 before then joining the Magpie group at Imperial College London for a post-doctoral research position. She conducted further research at the Lawrence Livermore National Laboratory, joining the team of scientists analysing data from the National Ignition Facility. She returned to Imperial College in April 2023 where she is continuing to work with collaborators and establish her own research team. 
\end{IEEEbiographynophoto}

\begin{IEEEbiographynophoto}{Jack Davies Hare,}
photograph and biography are not available at the time of publication.
\end{IEEEbiographynophoto}

\begin{IEEEbiographynophoto}{Jeremy P. Chittenden,}
photograph and biography are not available at the time of publication.
\end{IEEEbiographynophoto}

\begin{IEEEbiographynophoto}{Mark E. Koepke}
received all of his degrees from University of Maryland (Ph.D. in 1984). He is a professor of physics at West Virginia Univ. Research regimes of interest span low-temp partially ionized, space-related, fusion, and high-energy-density lab plasmas, as well as diagnostic methods and nonlinear dynamics.
\end{IEEEbiographynophoto}

\begin{IEEEbiographynophoto}{Eric G. Blackman}
is a Professor of Physics and Astronomy at the University of Rochester. He obtained undergraduate degrees in physics and math from MIT, a Master of Advanced Study in (applied) mathematics (Tripos Part III) at Cambridge University, and Ph.D. at Harvard University in theoretical astrophysics. He was a postdoc at the Institute of Astronomy, of Cambridge University and in physics at Caltech before joining U. Rochester in 2000. He has been a Fellow of the American Physical Society since 2004.
\end{IEEEbiographynophoto}

\begin{IEEEbiographynophoto}{Sergey V. Lebedev,}
photograph and biography are not available at the time of publication.
\end{IEEEbiographynophoto}






\end{document}